\documentclass[12pt]{iopart}

\usepackage{graphicx,mathrsfs}   %amsmath,
\usepackage{amssymb}
\usepackage{amsthm}
\usepackage{bm}
\usepackage{subfigure,xcolor}   % mathbbol   for math-1 etc....

\def\bc{\begin{center}}
\def\ec{\end{center}}

\newcommand{\bss}[1]{\boldsymbol{#1}}

\newcommand{\nn}{\nonumber}

\def\ie{\emph{i.e.},\ }
\def\eg{\emph{e.g.}\ }

\def\cf{\emph{c.f.\ }}
\newcommand{\pd}{{\phantom{\dag}}}
\newcommand{\up}{\uparrow}
\newcommand{\dw}{\downarrow}
\newcommand{\eps}{\varepsilon}
\begin{document}
\title[Quantum phase transitions of topological insulators without gap closing]{Quantum phase transitions of topological insulators without gap closing}
\author{Stephan Rachel}
\address{Institute for Theoretical Physics, TU Dresden, 01062 Dresden, Germany}
\ead{stephan.rachel@tu-dresden.de}

\begin{abstract}
We consider  two-dimensional Chern insulators and time-reversal invariant topological insulators 
and discuss the effect of perturbations breaking either particle-number conservation or time-reversal symmetry.
The appearance of trivial mass terms is expected to cause quantum phase transitions into trivial phases when such a perturbation overweights the topological term. These phase transitions are usually associated with a bulk-gap closing.
In contrast, the chiral Chern insulator is unaffected by particle-number breaking perturbations. Moreover, the $\mathbb{Z}_2$ topological insulator undergoes phase transitions into topologically trivial phases without bulk-gap closing in the presence of any of such perturbations. In certain cases, these phase transitions can be circumvented and the protection restored by another U(1) symmetry, \eg due to spin conservation.
These findings are discussed in the context of interacting topological insulators.
\end{abstract}
%\keywords{topological insulator, quantum phase transition}
\submitto{\JPCM}
\maketitle

\section{Introduction}

Quantum states of matter with bulk energy gap can be distinguished as phases with long-range entanglement and phases with short-range entanglement\,\cite{chen-10prb155138}. Fractional quantum Hall systems\,\cite{laughlin83prl1395}, chiral spin liquids\,\cite{kalmeyer-87prl2095} and $\mathbb{Z}_2$ spin liquids\,\cite{read-91prl1773} are examples of the first case, also referred to as phases with ``intrinsic topological order''; generally they do not require any symmetry. In contrast, short-ranged entangled phases rely on symmetries; once all symmetries are absent these states can be adiabatically connected with the same trivial product state.
If symmetries are present  short-range entangled phases can be distinguished. Such phases are called symmetry protected topological  (SPT) phases \cite{Volovik03,chen-10prb155138,pollmann-10prb064439,chen-11prb235128,turner-11prb075102,wen12prb085103,chen-12s1604,sule-13arXiv:1305.0700,gu-13arXiv:1304.4569}, prominent examples are the spin-1 Haldane chain\,\cite{haldane83prl1153} or $\mathbb{Z}_2$ topological insulators (TIs)\cite{kane-05prl146802,kane-05prl226801,bernevig-06s1757,koenig-07s766}. 
The concept of SPT phases has consequently been applied to classify quantum phases. 
In one spatial dimension, SPT phases were claimed to be completely classified by the elements in the second group cohomology class\,\cite{chen-11prb235128,turner-11prb075102}.
Also for non-interacting fermionic systems in $d$ spatial dimensions a complete characterization has been derived based on both group cohomology  and $K$-theory\,\cite{schnyder-08prb195125,kitaev08,ryu-10njp065010,wen12prb085103}. Classification based on SPT phases for interacting bosonic states of matter in $d>1$ are currently explored\,\cite{chen-12s1604}.
Recently, it has been proposed that long-ranged entangled phases can further be {\it enriched} by symmetries leading to the concept of ``symmetry enriched topological phases''\,\cite{mesaros-13prb55115,lu-16prb155121,huang-14prb045142}.

In this paper, we consider two-dimensional $\mathbb{Z}$ TIs, \ie Chern insulators, and $\mathbb{Z}_2$ time-reversal invariant TI models and study the effect of several symmetry-breaking terms. In the first part, we consider Chern insulators and study the effect of superconducting perturbations (\ie terms breaking particle-number conservation) which partially gives an answer to whether a Chern insulator is an SPT phase. 
In the second part, we focus on the $\mathbb{Z}_2$ TI which is an SPT phase protected by time-reversal symmetry and particle-number conservation. These non-interacting fermion models are 
characterized by a $\mathbb{Z}_2$ invariant and a Kramers pair of gapless helical edge states (per edge).
As mentioned above, when an SPT phase hast lost all its symmetries it can be adiabatically connected with topologically trivial band-insulator states, \ie simple product states. In the presence of symmetries, however, one expects a closing of the bulk gap at the phase transition. 
%In this picture, one can easily understand in which cases a quantum phase transition from a TI phase into another (possibly topologically trivial) phase is associated with closing of the bulk gap. 

Recently, topological quantum phase transitions (QPTs) without closing of the bulk gap have attracted considerable interest\,\cite{ezawa-13arXiv:1307.7347}. Also for interacting TI models the phase transition into the magnetically ordered phase is in some cases known to be associated with no bulk gap closing\,\cite{hohenadler-11prl100403,wu-12prb205102}. Moreover, the possibility of a TR symmetry broken TI phase, dubbed {\it spin-Chern} insulator, has been proposed recently\,\cite{ezawa-13arXiv:1307.7347,yoshida-13prb085134,miyakoshi-13arXiv:1304.7933}. 
%
%By using our insights about Chern insulators, we are able to introduce another spin-Chern insulator state where the U(1) particle-number is broken.

This paper is organized as follows:
In Sec.\ 2 we discuss a simple Chern insulator model in the presence of a  perturbation violating U(1) particle number conservation. In Sec.\ 3 we briefly discuss two paradigmatic TI models, the Kane-Mele and the Sodium-Iridate model. We discuss the effect of time-reversal breaking perturbations and demonstrate the differences between both models. Sec.\ 4 discusses the stability of the spin-Chern insulator phase. All these findings are in Sec.\ 5 contrasted with correlated TI models where the previously discussed situations occur within a mean-field treatment. We conclude in Sec.\ 6.

\section{Chern insulators}
Chern insulators or $\mathbb{Z}$ topological insulators are often defined as translation invariant band insulators featuring a finite (and quantized) Chern number $C$ which relates to the Hall conductance  via $\sigma^{xy}=\frac{e^2}{h}C$\,\cite{bernevig13}. Chern insulators realize the quantum Hall effect (QHE) and can be thus described as free fermion theories. By definition they are short-ranged entangled. The introduction might give the impression that Chern insulators must, hence, belong to the family of SPT phases. When the U(1) charge conservation symmetry (denoted as U(1)$_{\rm charge}$ in the following) is broken, \eg due to  superconducting proximity effect, the Hall conductance might loose its quantized value in general. Thus one could naively assume that a Chern insulator is an SPT phase protected by U(1)$_{\rm charge}$ symmetry.

\begin{figure}[h!]
\centering
\includegraphics[scale=0.6]{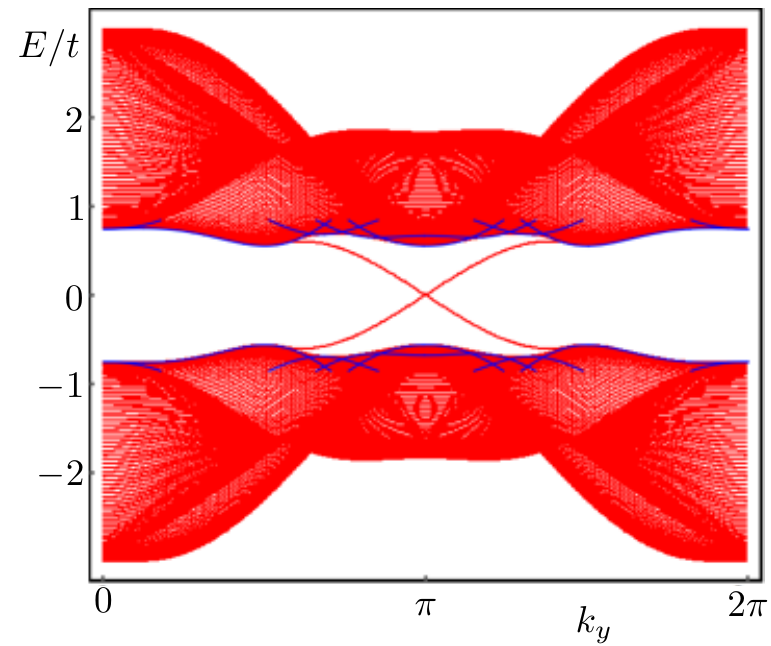} %{ci+NNpairing.png}
\caption{Spectrum of the Chern insulator model plus nearest-neighbor pairing, $\mathcal{H}_{\rm CI}+\mathcal{H}_2$, on a cylinder geometry. Parameters are $m=0.75\,t$, $\Delta_2=0.25\,t$, and the ribbon length corresponds to 64 unit cells. The number of edge modes and bands is doubled due to the redundancy of the Bogoliubov-de Gennes description. In blue, the bands closest to the gap obtained for PBCs are shown on order to demonstrate the agreement between ribbon and PBC calculation.}
\label{fig:ci+NNpairing}
\end{figure}

In the following, we explicitly test these statements and consider a Chern insulator model on the square lattice with additional BCS pairing term. The most simple pairing term is onsite $s$-wave,
\begin{equation}\label{H1}
\mathcal{H}_1 = \sum_i \left( \Delta_1 c_{i\up}^\dag c_{i\dw}^\dag + \Delta_1^\ast c_{i\dw}^\pd c_{i\up}^\pd \right)\ ,
\end{equation}
which requires spin. For a spinless fermion Chern insulator model, the simplest pairing is then a nearest-neighbor $s$-wave pairing of the form
%\begin{equation}\label{H2}
$\mathcal{H}_2' = \sum_{\langle ij \rangle} \left( \Delta_2 \, c_i^\dag c_j^\dag + \Delta_2^\ast \, c_j^\pd c_i^\pd \right)
$. %\end{equation}
As a toy model we consider the  Chern insulator model regularized on a square lattice\,\cite{QiHughesZhang2009} which corresponds to a single spin-channel of the Bernevig--Hughes--Zhang model\,\cite{bernevig-06s1757},
\begin{equation}\label{ham:ci}
\mathcal{H}_{\rm CI} = \sum_{\boldsymbol{k}} \left( c_{s,\boldsymbol{k}}^\dag ~c_{p,\boldsymbol{k}}^\dag \right) h(\boldsymbol{k}) \left(\begin{array}{c}c_{s,\boldsymbol{k}}\\[5pt] c_{p,\boldsymbol{k}}\end{array}\right)\ ,
\end{equation}
where $h(\boldsymbol{k})=\boldsymbol{d}\cdot \boldsymbol{\sigma}$ with $d_x = t\sin{k_x}$, $d_y=t\sin{k_y}$, $d_z = m + \cos{k_x} + \cos{k_y}$. The Pauli matrices $\boldsymbol{\sigma}$ correspond to an orbital subspace, say, spanned by $s$ and $p$ orbitals. We add the $\mathcal{H}_2'$ BCS term in this basis,
\begin{equation}\label{H2}
\mathcal{H}_2 = \sum_{\langle ij \rangle}\sum_{\alpha=s,p} \left( \Delta_2 \, c_{\alpha,i}^\dag c_{\alpha,j}^\dag + \Delta_2^\ast \, c_{\alpha,j}^\pd c_{\alpha,i}^\pd \right)\ .
\end{equation}
We compute the energy spectrum of $\mathcal{H}=\mathcal{H}_{\rm CI}+\mathcal{H}_2$ on a cylinder having the benefit that we can easily keep track of the stability of the Chern insulator phase as signaled by its chiral edge modes. See Fig.\,\ref{fig:ci+NNpairing} for an example.
Note that the right-moving (left-moving) edge mode is located at the left (right) edge of the cylinder.
%The beforementioned doubling of energy bands also affects the edge modes; to compensate for that only the edge states with $E/t \geq 0$ are meaningful.

%%%%%%%%%%%%%%%%%%%%%%%%%%%%%%%%%%%%%%%
\begin{figure}[t!]
\centering
\includegraphics[scale=0.5]{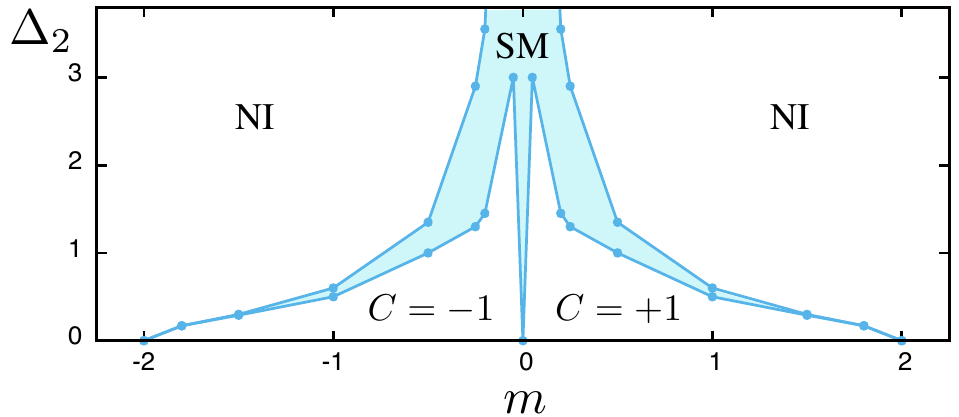}
\caption{$m$-$\Delta_2$ phase diagram of $\mathcal{H}_{\rm CI}+\mathcal{H}_2$ containing normal insulating (NI) and semi-metallic (SM) phases as well as the topological Chern insulator phases with $C=\pm 1$.}
\label{fig:m-Delta2}
\end{figure}

%%%%%%%%%%%%%%%%%%%%%%%%%%%%%%%%%%%%%%%
Note that the $\mathcal{H}_2$ term forces us to reformulate the Bloch matrix $h(\boldsymbol{k})$ in the new Bogoliubov--de Gennes basis as usual. This causes doubling of the energy bands or, with other words, each energy level at $+\eps$ also exists at energy $-\eps$. We find that finite $\Delta_2$ does not gap out the chiral edge modes (even when $\Delta_2/t\approx 1$, see the $m$-$\Delta_2$ phase diagram, Fig.\,\ref{fig:m-Delta2}). That is, the Chern insulator  is stable in the presence of a U(1)$_{\rm charge}$ breaking perturbation.

%{\color{red} 
The phase diagram Fig.\,\ref{fig:m-Delta2} hosts the Chern insulator phase ($C=\pm 1$), a normal-insulating (NI) phase (\ie topologically trivial) and a gapless Dirac semi-metal phase. With other words, also in the presence of $\Delta_2$ the bulk gap must close when the Chern number $C$ changes. Note that for finite $\Delta_2$ the system is superconducting (for instance, induced by proximity to a superconductor) -- instead of ``normal insulating'' one could also call this phase ``trivial superconducting'' etc.
%}

%
%
We interpret these findings as follows. Upon breaking the U(1)$_{\rm charge}$ symmetry a ``charge'' Hall conductivity cannot  remain quantized in general\,\cite{volovik88zetf123}. If one defines the Chern insulator as a band insulator with quantized $\sigma^{xy}$, a BCS pairing term trivially destroys such a phase. In the context of SPT phases, the given definition is, however, misleading. Instead one should rather define the Chern insulator as a phase with a finite (and quantized) Chern number $C\in\mathbb{Z}\backslash\{0\}$ as we did previously. The number $C$ corresponds to the number of chiral edge modes per edge when a disk or cylinder geometry is considered, respectively. While $\sigma^{xy}$ might loose its quantized value and $\sigma^{xy} \not= \frac{e^2}{h}C$ in general, $C$ itself remains unaffected as demonstrated in Fig.\,\ref{fig:ci+NNpairing}. 
%%%XXX
Note that the part of the Hall conductance related to the edge states is in topological superconductors generally expected to be quantized\,\cite{volovik92jetp368}.
The Chern insulator is not an SPT phase protected by U(1)$_{\rm charge}$ symmetry. In fact, a Chern insulator (and the QHE) is neither topologically ordered (or long-ranged entangled) nor symmetry-protected. It is simply a chiral phase due to broken time-reversal symmetry.
%%%%%%%%%%%%%%%%%%%%%%%%%%%%%%%%%%%%%%%
%%%%%%%%%%%%%%%%%%%%%%%%%%%%%%%%%%%%%%%

%{\color{red}
Let us finish this section with a brief discussion of topological classes.
In Refs.\,\cite{schnyder-08prb195125,ryu-10njp065010} all possible phases of non-interacting fermions have been characterized solely based on their behavior under time-reversal ($\mathcal{T}$) and charge-conjugation ($\mathcal{C}$) symmetry. These symmetries are special as they correspond to anti-unitary operators. There are three possibilities for TR symmetry: (i) absence ($\mathcal{T}=0$) or (ii) presence with the squared TR operator being $+1$ ($\mathcal{T}=+1$) or (iii) presence with the squared TR operator being $-1$ ($\mathcal{T}=-1$). Then there are the three analogous cases for charge-conjugation symmetry, $\mathcal{C}=0, \pm1$. This scheme gives rise to $3\times 3=9$ different classes. In addition, the chiral symmetry $\mathcal{S}=\mathcal{T}\cdot\mathcal{C}$ is considered ($\mathcal{S}=0$ or 1). For eight of these nine cases $\mathcal{S}$ is fixed, only one case is split into two different classes. In total, there are ten symmetry classes, which are the same as the Cartan classes and the classification scheme corresponds to the one of Altland and Zirnbauer\,\cite{zirnbauer96jmp4986,altland-97prb1142}. The time-reversal invariant topological insulators in $d=2$ (later discussed in this paper) and in $d=3$ are in symmetry class AII (with $\mathbb{Z}_2$ invariant). The integer QHE and Chern insulators are in the class where all symmetries are absent, class A (with $\mathbb{Z}$ invariant, the Chern number). Upon adding a BCS term and switching from a Bloch matrix description to a Bogouliubov--de Gennes matrix description, the system clearly becomes particle-hole symmetric, \ie $\mathcal{C}=+1$. Thus the symmetry class changes from A to D (which still has a $\mathbb{Z}$ invariant).
%}

\section{Topological insulators}

\begin{figure}[t!]
\centering
\includegraphics[scale=0.6]{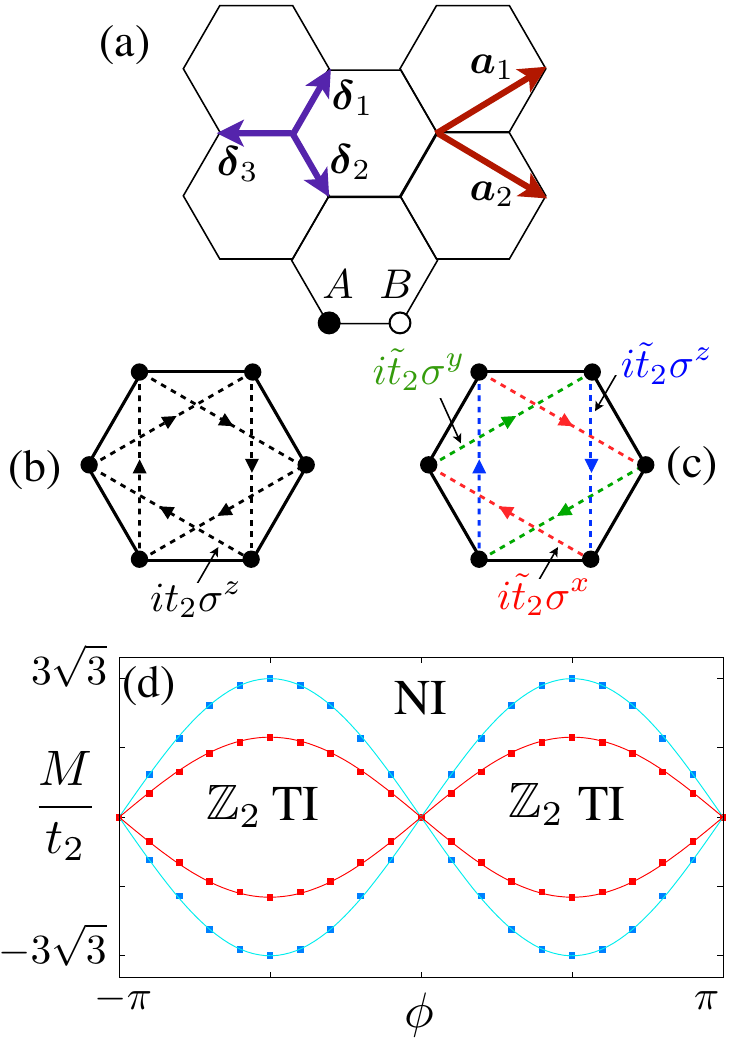}
\caption{(Color online). (a) Honeycomb lattice with nearest neighbor vectors $\boldsymbol{\delta}_i$, translation vectors $\boldsymbol{a}_i$, and sublattices $A$, $B$. (b) Second neighbor spin-orbit hoppings of the KM model and (c) of the SI model. (d) $M$-$\phi$ phase diagrams ($t_2\equiv\tilde t_2$) of (i) KM model (blue curve) and (ii) SI model  (red curve) with topological insulator (TI) and normal insulator (NI) phases. Data points correspond to the gap-closing transitions obtained numerically and the solid lines are derived from the continuum models, see (\ref{km}) and (\ref{si}).}
\label{fig:para-km+shit}
\end{figure}

Now we consider time-reversal (TR) invariant $\mathbb{Z}_2$ TIs in two spatial dimensions. We start with the honeycomb TI models and consider both the celebrated Kane-Mele (KM) model\,\cite{kane-05prl146802,kane-05prl226801} as well as sodium iridate (SI) model\,\cite{shitade-09prl256403} which is a bond-dependent generalization of the KM model. The major difference is that the KM model in the absence of Rashba spin-orbit coupling preserves the axial spin symmetry (\ie U(1)$_{\rm spin}$) while the SI model breaks the U(1)$_{\rm spin}$ symmetry.
The KM model is governed by the Hamiltonian
\begin{eqnarray}\nn
&&\mathcal{H}_{\rm KM} \!=\! -t \sum_{\langle ij \rangle} c_{i\sigma}^\dag c_{j\sigma}^\pd 
\!+ i t_2 \sin{\phi}
\sum_{\langle\!\langle ij \rangle\!\rangle} 
%\sum_{\alpha\beta} 
\nu_{ij} c_{i\alpha}^\dag \sigma^z_{\alpha\beta} c_{j\beta}^\pd \\[5pt]
\nn&&\!+\! t_2 \cos{\phi}\sum_{\langle\!\langle ij \rangle\!\rangle\,}  c_{i\sigma}^\dag c_{j\sigma}^\pd
  + i\lambda_R \sum_{\langle ij \rangle} 
%\sum_{\alpha\beta} 
c_{i\alpha}^\dag (\boldsymbol{\sigma}_{\alpha\beta} \times \boldsymbol{d})_z c_{j\beta}^\pd\
\end{eqnarray}
where we assume summation over spin indices $\sigma$, $\alpha$, and $\beta$. The first line contains the minimal KM model, real spin-independent nearest neighbor hopping and imaginary spin-dependent second neighbor hopping (\ie spin-orbit coupling) breaking the SU(2) symmetry down to U(1) spin symmetry, \ie $S^z$ still is conserved. The second line contains a real spin-independent second neighbor hopping breaking particle-hole symmetry and a Rashba spin-orbit term breaking both the U(1) spin symmetry and the $z\to-z$ mirror symmetry. $\nu_{ij}=(\boldsymbol{d}_{kj}\times\boldsymbol{d}_{ik})/|\boldsymbol{d}_{kj}\times\boldsymbol{d}_{ik}|=\pm 1$ and the $\boldsymbol{d}_{(ij)}$ are the corresponding nearest-neighbor vectors $\boldsymbol{\delta}_\mu$ ($\mu=1,2,3$) depending on which bonds are involved when hopping from $i$ to $j$, see Fig.\,\ref{fig:para-km+shit}\,(a).
%In contrast to Refs.\,\onlinecite{kane-05prl146802,kane-05prl226801} 
Here we consider arbitrary $\phi$ in the spirit of Haldane's seminal paper from 1988\,\cite{haldane88prl2015}. The related SI model\,\cite{shitade-09prl256403} is described by the Hamiltonian
\begin{eqnarray}
\mathcal{H}_{\rm SI} &=& -t \sum_{\langle ij \rangle} c_{i\sigma}^\dag c_{j\sigma}^\pd + \tilde t_2 \cos{\phi}\sum_{\langle\!\langle ij \rangle\!\rangle\,}  c_{i\sigma}^\dag c_{j\sigma}^\pd\\[5pt]
 &&+ i \tilde t_2 \sin{\phi}
\sum_{\langle\!\langle ij \rangle\!\rangle_\gamma} 
%\sum_{\alpha\beta} 
\nu_{ij} c_{i\alpha}^\dag \sigma^\gamma_{\alpha\beta} c_{j\beta}^\pd\ .
\end{eqnarray}
The SI model can be seen as an extension of the KM model: it possesses a multi-directional, bond-dependent spin orbit coupling. For the second-neighbor links in vertical direction ($\gamma=z$) it corresponds to $\pm i \tilde t_2 \sigma^z$, in $\boldsymbol{a}_2$ direction ($\gamma=x$) to $\pm i \tilde t_2 \sigma^x$, and in $\boldsymbol{a}_1$ direction ($\gamma=y$) to $\pm i \tilde t_2 \sigma^y$. This convention of spin-orbit hopping is illustrated in Fig.\,\ref{fig:para-km+shit}\,(b) for the KM and in Fig.\,\ref{fig:para-km+shit}\,(c) for the SI model.
As long as the second neighbor hopping is complex ($\phi\not=0, \pm\pi$) both KM and SI models are $\mathbb{Z}_2$ topological insulators (at least when $\lambda_R \leq t_2$) and belong to the same universality class. That is, the interpolating Hamiltonian $H(\alpha)=(1-\alpha)\mathcal{H}_{\rm KM} + \alpha \mathcal{H}_{\rm SI}$ remains gapped and the helical edge states persist for $\alpha\in [0,1]$ (for $\phi\not=0, \pi$, $0<t_2, \tilde t_2<t$  and $\lambda_R=0$). Let us emphasize once more that the spin orbit term of the SI model breaks the axial spin symmetry while the spin orbit term  of the KM model still preserves $S^z$. In the Appendix, an explicit momentum-space representation of $\mathcal{H}_{\rm KM}$ and $\mathcal{H}_{\rm SI}$ is presented.
%
%%%%%%%%%%%%%%%%%%%%%%%%%%%%%%%%%%%%%%%
%%%%%%%%%%%%%%%%%%%%%%%%%%%%%%%%%%%%%%%

\subsection{Quantum phase transitions with gap-closing}

We first consider simple cases of topological QPTs. On the honeycomb lattice, the most natural term to drive the system into a topologically trivial phase is the Semenoff mass term\,\cite{semenoff84prl2449} (aka {\it staggered potential}) as already pointed out by Haldane\,\cite{haldane88prl2015}. It is given by
\begin{equation}
\mathcal{H}_S = M \sum_{i\sigma} \left( a_{i\sigma}^\dag a_{i\sigma}^\pd - b_{i\sigma}^\dag b_{i\sigma}^\pd \right)\ ,
\end{equation}
and the electron operator $a_{i\sigma}$ ($b_{i\sigma}$) annihilates an electron on sublattice A (B), see Fig.\,\ref{fig:para-km+shit}\,(a). 
For $\lambda_R=0$, the KM model consists of two decoupled copies of Haldane's model, thus the $M$-$\phi$ phase diagram of the KM model is identical to the one by Haldane\,\cite{haldane88prl2015}. 
The $M/t_2$-$\phi$ phase diagram of the KM model (SI model) is shown as the blue (red) curve in Fig.\,\ref{fig:para-km+shit}\,(d) containing TI and normal insulating (NI) phases, the blue (red) lines denote the gap-closing transition (as a Dirac semi-metal)  between TI and NI phases.

For the Haldane model (and thus for the KM model) the phase diagram can be obtained analytically by considering the continuum model. Since gap-closing occurs only at the corners of the first Brillouin zone $\boldsymbol{K}$ and/or $\boldsymbol{K}'$, respectively, one might expand  the Bloch matrix around these points. The terms on the diagonal being proportional to $\eta^z$ (Pauli matrix associated with the sublattice)  correspond to the competing mass terms (Semenoff term vs.\ Haldane term) of the Dirac-Hamiltonian. The gap-closing condition is thus 
\begin{equation}\label{km}
M/t_2 = \pm 3\sqrt{3} \sin{\phi}\ .
\end{equation}
For the SI model one obtains the similar result
\begin{equation}\label{si}
M/{\tilde t_2} = \pm 3 \sin{\phi}
\end{equation}
in agreement with the numerical detection of the gap-closing points, see Fig.\,\ref{fig:para-km+shit}\,(d).
Eq.\,(\ref{km}) can be directly read off from the diagonal of the Bloch matrix at momentum $\bss{k}=\bss{K}$; in order to obtain Eq.\,(\ref{si}) one needs to diagonalize the Bloch matrix at $\bss{k}=\bss{K}$. It reads $h_{\rm SI}(\bss{K}) = $
\begin{equation*}
\small
\left(\!\!\!\begin{array}{cccc}
M+\sqrt{3} \tilde t_2 \sin{\phi} & 0 & (1-i)\sqrt{3} \tilde t_2 \sin{\phi} & 0 \\
0 & -M- \sqrt{3} \tilde t_2 \sin{\phi} & 0 & (-1+i)\sqrt{3}\tilde t_2 \sin{\phi} \\
(1+i) \sqrt{3}\tilde t_2 \sin{\phi} & 0 & M - \sqrt{3} \tilde t_2 \sin{\phi} & 0 \\
0 & (-1-i)\sqrt{3} \tilde t_2 \sin{\phi} & 0 & -M+\sqrt{3}\tilde t_2\sin{\phi}
\end{array}\!\!\!\right)
\end{equation*}
and possesses the eigenvalues $\pm(M\pm 3\tilde t_2 \sin{\phi})$ leading to (\ref{si}).

The Semenoff mass term does not provide the only way to drive the system into a topologically trivial phase. Lattice anisotropies such as plaquette anisotropy\,\cite{wu-12prb205102} or $t_1$-$t_2$ anisotropy\,\cite{lang-13prb205101} as well as real third-neighbor hopping\,\cite{hung-13prb121113,hung-13arXiv:1307.2659} might cause a transition into a trivial band insulator phase, just to mention a few. Recently, also the phase transitions (associated with gap-closing) between spin-orbit  dominated TI phase and  magnetic field dominated quantum Hall phase have been investigated\,\cite{goldman-12epl23003,beugeling-12prb075118}.

\subsection{Quantum phase transitions without gap-closing}

In contrast to the previous section, we consider now time-reversal (TR) breaking masses. Having the naive picture of SPT phases in mind we expect that an applied TR breaking term immediately destroys the topological phase since the $\mathbb{Z}_2$ TI is protected by TR symmetry. We consider two examples: (i) the KM model without Rashba spin orbit coupling but with in-plane staggered magnetization. This scenario corresponds to the proper mean-field treatment of the Kane-Mele-Hubbard model\,\cite{rachel-10prb075106,hohenadler-11prl100403,wu-12prb205102,soriano-10prb161302,dong-11prb205121,yamaji-11prb205122,yu-11prl010401,lee11prl166806,hohenadler-12prb115132,griset-12prb045123,hohenadler-13jpcm143201,araki-13prb205440}. (ii) The SI model with staggered magnetization pointing in arbitrary direction. 

Since infinitesimally small magnetization changes the phase immediately the term ``quantum phase transition" might be misleading; but this considerations are motivated by the static mean-field analysis of such models with additional Hubbard onsite interaction with amplitude $U$. In this mean-field picture, a critical $U_c$ is required to induce a finite magnetization and thus the term ``quantum phase transition'' is clearly justified.

\subsubsection{Kane-Mele model with in-plane staggered magnetization\\[5pt]}

We apply an antiferromagnetic in-plane Zeeman field $m_{\rm af} (S_A^x - S_B^x)$ where the index $A$ ($B$) refers to sublattice $A$ ($B$) of the honeycomb lattice. The spin operator can be expressed by fermion operators, $2 S^x=S^++S^- = c_{\up}^\dag c_{\dw}^\pd + c_{\dw}^\dag c_{\up}^\pd$. Thus in the Bloch basis $\Psi_{\boldsymbol{k}}^\dag=(a_{\boldsymbol{k}\up}^\dag, b_{\boldsymbol{k}\up}^\dag, a_{\boldsymbol{k}\dw}^\dag, b_{\boldsymbol{k}\dw}^\dag )$ this term has the following form,
\begin{equation}
S_A^x - S_B^x = \frac{1}{2}\left(\begin{array}{cccc}
&&1&0\\
&&0&-1\\
1&0&&\\
0&-1&&
\end{array}\right)\ .
\end{equation}
As expected from an SPT phase, we find immediate gapping of the edge states (indicating the loss of the protecting TR symmetry) while the bulk gap stays open (\cf Ref.\,\cite{timm-12prb155456}). Now one could change another band structure parameter to tune the system into the atomic trivial limit. Note that the right choice of such a parameter is not always obvious. The statement is only about the existence of such a path in parameter space once  the relevant symmetries are broken.

\subsubsection{Sodium iridate model with arbitrary staggered magnetization\\[5pt]}

The second example we consider is the SI model with additional antiferromagnetic Zeeman field. We find immediate gapping of the edge states for arbitrary direction of Zeeman field.
This result is consistent with the previous findings: since the spin-orbit coupling fully breaks the axial spin symmetry, any TR breaking field breaks the topological protection. From Fig.\,\ref{fig:para-km+shit}\,(c) one sees that Zeeman fields proportional to any Pauli matrix $\sigma^i$ should have the same effect in agreement with the study of energy spectra.

Both systems considered here are examples of topological QPTs {\it without} gap-closing. We stress again that this is by no means unexpected since the phase transition is associated with getting rid of the protecting symmetry of the topological phase. In contrast, this is simply a confirmation of what we already learned from the first TI papers. Note that similar situations would occur by applying an U(1)$_{\rm charge}$ symmetry breaking term such as (\ref{H1}). Furthermore, we emphasize that both considered scenarios correspond to the mean-field descriptions of the KM-Hubbard (KMH) model and SI-Hubbard (SIH) model: the KMH model exhibits an easy plane antiferromagnetic phase for large Hubbard-$U$\,\cite{rachel-10prb075106,hohenadler-11prl100403,dong-11prb205121,wu-12prb205102} while the SIH model is Neel ordered for large $U$ (at least as long as $|t_2|\ll |t|$)\,\cite{reuther-12prb155127,ruegg-12prl046401,kargarian-12prb205124,liu-13arXiv:1307.4597,rachel-15prl167201}.

\section{Spin-Chern insulator}
\subsection{TR-breaking spin-Chern insulator}
\begin{figure}[t!]
\centering
\includegraphics[scale=0.6]{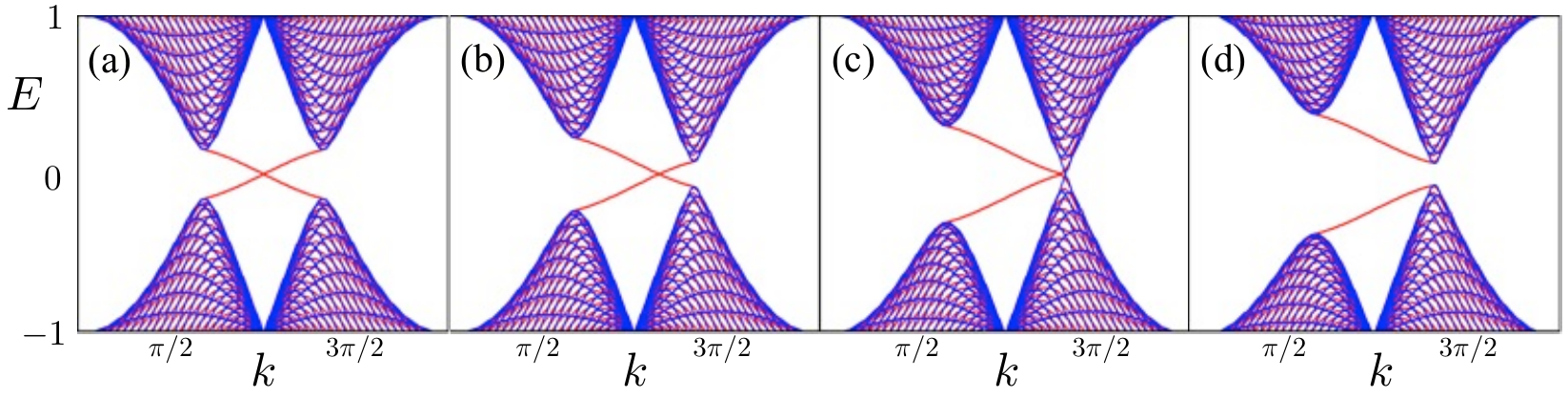}
\caption{Spectra of the KM model ($t_2=0.03\,t$, $\lambda_R\equiv 0$) for various values of antiferromagnetic Zeeman field $m_{\rm af} (S_A^z - S_B^z)$. (a) TI phase for $m_{\rm af}=0$, (b) spin-Chern insulator for $m_{\rm af}=0.15\,t$,  (c) semi-metal for $m_{\rm af}=0.3\,t$, and (d) trivial insulator for $m_{\rm af}=0.45\,t$. The band structure for a cylinder (red) is superimposed by the band structure for a torus (blue).}
\label{fig:km+maz-edge}
\end{figure}

Now we turn to the interesting case where one applies an antiferromagnetic Zeeman field to the KM model with a magnetization pointing in the $z$-direction, $m_{\rm af}(S_A^z-S_B^z)$. This choice of Zeeman field explicitly contains the operator $S^z$, thus the axial spin symmetry remains intact and $S^z$ a good quantum number (with other words, this Zeeman term commutes with the Hamiltonian as long as $\lambda_R=0$). Although the Zeeman field breaks TR symmetry, the topological phase remains stable and the edge states persist. This phase is merely protected by the axial spin symmetry and usually referred to as {\it spin-Chern insulator}\,\cite{ezawa-13arXiv:1307.7347}. Since spin is a good quantum number, Chern numbers for $\up$- and $\dw$-spin channel can be calculated independently and the spin Chern number $C_S = C_\up - C_\dw$ is a well-defined quantity. Further increase of the Zeeman field eventually causes a gap-closing at the $\boldsymbol{K}$- or $\boldsymbol{K}'$-point in the Brillouin zone, respectively, before the system enters a topologically trivial phase. In Fig.\,\ref{fig:km+maz-edge} we have plotted spectra which illustrate the different scenarios. In panel (a) we show the pure KM model for $m_{\rm af}=0$. In panel (b), we show the spectrum of a spin Chern insulator with its edge states for $m_{\rm af}=0.15\,t$. The $\eps(k)\not= \eps(-k)$ asymmetry reflects the broken TR symmetry. The panel (c) shows the gap-closing as a Dirac theory at the $\boldsymbol{K}$-point ($m_{\rm af}=0.3\,t$) and eventually in panel (d) the topologically trivial phase is shown.
%
%%%%%%%%%%%%%%%%%%%%%%%%%%%%%%%%%%%%%%%%%%%%%%%%
%%%%%%%%%%%%%%%%%%%%%%%%%%%%%%%%%%%%%%%%%%%%%%%%
%
%{\color{red}
Note that on each edge of the considered cylinder a pair of helical edge states is present. The edge states crossing the bulk gap are thus degenerate. The right (left) mover at the left edge is a spin-up (spin-down) mode and vice versa at the other edge.
%}

\begin{figure}[t!]
\centering
\includegraphics[scale=0.65]{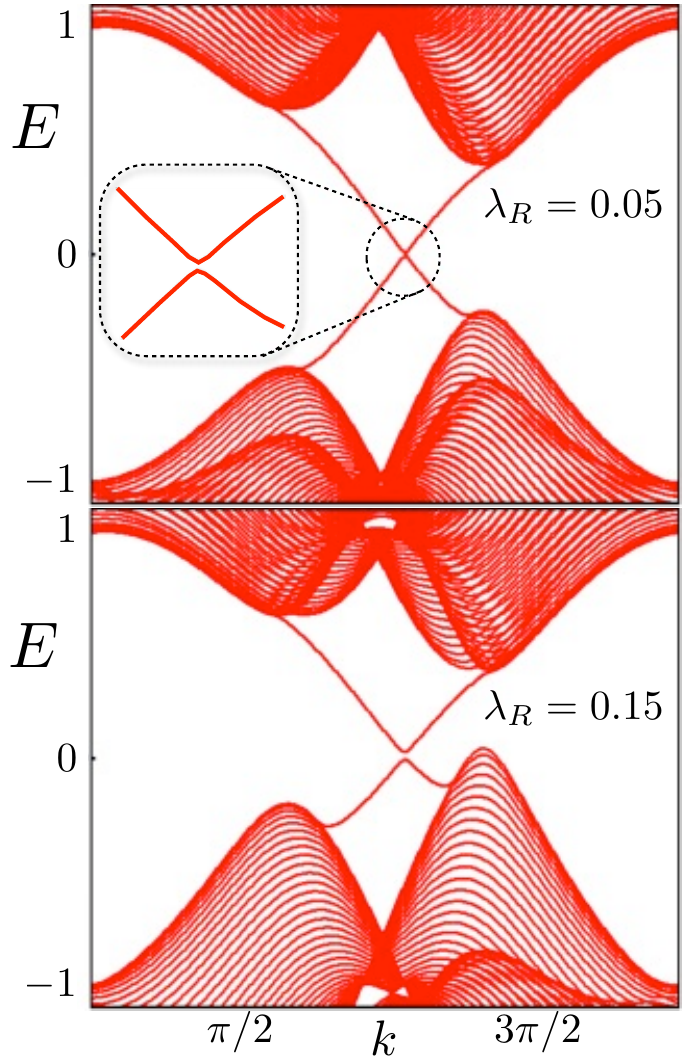}
\caption{Spectrum of the spin-Chern insulator  which is  destroyed by the immediate presence of Rashba spin orbit coupling. Parameters are $t_2=0.1\,t~ (\phi=\pi/2)$, $m_{\rm af}=0.25\,t$, and $M=0$ on a ribbon with 48 unit cells: (top) $\lambda_R=0.05\,t$; (bottom) $\lambda_R=0.15\,t$.}
\label{fig:kmr_gapped_example}
\end{figure}

If our consideration of the previous sections is meaningful we should test the stability of the spin-Chern insulator. The Semenoff mass term does not  affect the spin-Chern insulator (unless it becomes too large, see Sec.\,II).
In contrast, a $S^z$-symmetry breaking term should destroy the spin-Chern insulator.  A realistic term which breaks the axial spin symmetry is the Rashba spin orbit coupling as induced by an external electric field or a substrate\,\cite{kane-05prl146802}. Even for tiny amplitude $\lambda_R$ we find gapping of the edge states. This is illustrated in Fig.\,\ref {fig:kmr_gapped_example} for a ribbon exhibiting 48 unit cells and parameters $t_2=0.1$ ($\phi=\pi/2$), $m_{\rm af}=0.25\,t$, and $\lambda_R=0.05\,t$ and $\lambda_R=0.15\,t$, respectively. To rule out that the gapping of edge states is just due to the finite size of the ribbon we verified it on ribbons with 96 and 150 unit cells. 
%{\color{red}
Note that the pair of left- and right-moving edge states crossing the bulk gap is degenerate because there is another pair at the other edge. An additional Semenoff mass term $M$ would lift this degeneracy\,\cite{kane-05prl146802}. In contrast to Fig.\,\ref{fig:km+maz-edge} here spin is not conserved anymore and the edge modes are superpositions of up- and down-spin electrons. Moreover, the spin quantization axis is not only tilted it even rotates as a function of momentum\,\cite{schmidt-12prl156402,rod-15prb245112}.
%}

We have seen that the spin-Chern insulator exists. It is very similar to the $\mathbb{Z}_2$ TI. While the later is protected by TR and U(1)$_{\rm charge}$ symmetries, the spin-Chern insulator is protected by U(1)$_{\rm spin}$ and U(1)$_{\rm charge}$ symmetries. As such the spin-Chern insulator phase turns out to be rather fragile. In condensed matter experiments it will be difficult to guarantee absence of any $S^z$ breaking sources (such as substrates or electric fields), but cold atomic gases with their high degree of tuneability might provide a promising setup to realize this spin-Chern insulator phase.

\section{BHZ Hubbard model}
The honeycomb lattice topological insulators are beautiful model systems but despite many promising proposals\,\cite{weeks-11prx021001,liu-11prl076802,ezawa12epl67001,ghaemi-12prb201406,tang-13arXiv:1307.8054,murakami06prl236805,xu-13arXiv:1306.3008,weng-13arXiv:1309.7529,kou-13arXiv:1310.2580,roy-13prb045425} they have not been unambiguously verified in experiments. Nonetheless amongst them are a few good candidate materials such as silicene\,\cite{ezawa12prl055502,rachel-15prb195303}. The first experimental realization of the 2D topological insulator succeeded in HgTe/CdTe quantum wells\,\cite{koenig-07s766} following an earlier proposal by Bernevig, Hughes, and Zhang (BHZ)\,\cite{bernevig-06s1757}. In the minimal version of this BHZ model describing the TI phase of the HgTe/CdTe quantum wells the spin is conserved as in the KM model. Additional spin-symmetry breaking terms such as bulk and structural inversion asymmetry\,\cite{bernevig-06s1757,rothe-10njp065012} have also been proposed.
One might regularize the BHZ model on a two-orbital square lattice and add onsite Hubbard interactions resulting in the BHZ Hubbard model\,\cite{yoshida-13prb085134,miyakoshi-13arXiv:1304.7933,tada-12prb165138,budich-13prb235104}.
For strong interactions, it was claimed, however, that the TI phase turns into an Ising antiferromagnet with magnetization in $z$-direction\,\cite{miyakoshi-13arXiv:1304.7933} which is in contrast to the Kane-Mele-Hubbard model where the magnetic phase exhibits an in-plane magnetization. As such, the BHZ Hubbard model provides all ingredients which are necessary for a spin-Chern insulator phase. Indeed, such a {\it antiferromagnetic topological insulator} phase was proposed recently\,\cite{yoshida-13prb085134,miyakoshi-13arXiv:1304.7933}  (corresponding to the spin-Chern insulator discussed earlier) and confirmed using dynamical mean-field theory (DMFT) and the variational cluster approach (VCA).  Both DMFT and VCA implement the symmetry breaking fields, so-called {\it Weiss fields}, explicitly, what is similar to our discussion with applied staggered Zeeman fields. Following our claim from the previous section, we expect that this phase is not stable with respect to any bulk inversion asymmetry\,\cite{bernevig-06s1757} or Rashba spin orbit coupling\,\cite{rothe-10njp065012} (since both break the axial spin symmetry). DMFT and VCA methods still inherit some mean-field-like character. Hence, it would be very interesting to perform numerically exact quantum Monte Carlo simulations for the BHZ Hubbard 
model\,\cite{3DBHZ}, investigate its magnetic properties, and to test whether or not the spin-Chern insulator phase exists. There are good reasons to assume that additional terms or anisotropies lead to an even richer phase diagram. In a recent study of the magnetic properties of a topological Hubbard model on the square lattice\,\cite{cocks-12prl205303,orth-13jpb134004} 
a plethora of conventional and exotic magnetic ground states have been found\,\cite{orth-13jpb134004}. It was also shown that its non-interacting low-energy theory corresponds to an anisotropic BHZ model\,\cite{scheurer-15sr8386}.

As already pointed out the analogous scenario in the Kane-Mele-Hubbard (KMH) model does not exist since an in-plane antiferromagnetic order is energetically preferred. Nonetheless let us perform  a Gedankenexperiment where a magnetization in $z$-direction is present. In that case, the spin-Chern insulator scenario would occur in the phase diagram of the KMH model (as long as $\lambda_R=0$). Using a numerical method such as VCA it can be easily shown that KMH model with such an Ising-type Weiss field pointing in the $z$-direction exhibits the following phase diagram: for  $U<U_{c,1}$ the topological insulator is stable, at $U_{c,1}$ the magnetization becomes finite but the edge states persist (despite broken TR symmetry) and the bulk gap remains finite. For a larger $U_{c,2}$ the bulk gap closes indicating the transition into the topologically trivial magnetic phase for $U>U_{c,2}$. Thus the phase for $U_{c,1}<U<U_{c,2}$ corresponds to the spin-Chern insulator. As emphasized above, this scenario is energetically unstable for the isotropic, unperturbed KMH model and the true ground state situation is associated with easy-plane antiferromagnetic order and absence of a spin-Chern insulator phase. 
%%%XXX
Finally we wish to remark that a first order phase transition is another possibility  to avoid the gap closing when passing from a (correlated) topological insulator phase into a topologically trivial Mott insulator phase\,\cite{amaricci-15prl185701}.

\section{Discussion and Conclusion}

When TR symmetry is broken and no other symmetry is available to ``replace'' TR symmetry the topological protection is lost indicated by the small gap in the edge state spectrum. Naively one might think that elastic single-electron backscattering is now allowed leading to a low-dissipation spin transport. 
Recently, a very similar situation was studied using noncommutative spin-Chern numbers in the presence of disorder\,\cite{xu-12prb075115,yang-11prl066602}. It was claimed, however, that the metallic bulk states which are connected with the edge states (bulk-boundary correspondence) residing away from the Fermi level seem to survive localization\,\cite{xu-12prb075115}. These results imply that not TR symmetry but topology alone protects these bulk extended states in the strong-disorder regime. With other words, breaking of TR symmetry in 2D TIs does not lead to sudden and complete localization\,\cite{xu-12prb075115}. Clearly, these findings in the presence of disorder need to be clarified in the context of  symmetry protected topological phases; it seems that the non-trivial topology still survives once all the symmetries are gone. This opens an interesting perspective for the study of transport in disordered topological phases and, in particular, calls for further investigation of the spin-Chern insulator phase. As long as one defines the topological phase by means of helical edge states only, the picture provided by the theory of symmetry protected topological phases describes the spin-Chern insulator scenario and its stability very well.

\section{Acknowledgements}
The author acknowledges  discussions with Motohiko Ezawa, Karyn Le Hur, Ronny Thomale, Manuel Laubach, Frank Pollmann, Carsten Timm and thanks Matthias Vojta for a stimulating discussion. SR is supported by the DFG through priority program SPP 1666 ``Topological Insulators'' and through SFB 1143 ``Correlated Magnetism''.

\appendix
\section{Bloch matrices of topological insulator models}

In the following the momentum space representations of both $\mathcal{H}_{\rm KM}$ and $\mathcal{H}_{\rm SI}$ are explicitly listed for $\phi=\pi/2$. Quite generally we can write $\mathcal{H}=\sum_{\boldsymbol{k}} \Psi^\dag_{\boldsymbol{k}} H(\boldsymbol{k}) \Psi^\pd_{\boldsymbol{k}}$ with the four-component spinor $\Psi_{\boldsymbol{k}}^\dag=(a_{\boldsymbol{k}\up}^\dag, b_{\boldsymbol{k}\up}^\dag, a_{\boldsymbol{k}\dw}^\dag, b_{\boldsymbol{k}\dw}^\dag )$. The Bloch matrix reads
\begin{equation}
H(\boldsymbol{k}) = \sum_\mu g_\mu \Gamma_\mu + \sum_{\mu\nu} g_{\mu\nu}\Gamma_{\mu\nu}\ .
\end{equation}
As usual, the $4\times 4$ matrices $\Gamma_1, \ldots, \Gamma_5$ form a Clifford algebra, $\{\Gamma_\mu, \Gamma_\nu\} = 2 \delta_{\mu\nu}$ ($\mu, \nu=1,\ldots, 5$); the other ten matrices are constructed by the commutators $\Gamma_{\mu\nu}=\frac{1}{2i}[\Gamma_\mu, \Gamma_\nu]$. The generators $\Gamma_\mu$ can be chosen as $\vec \Gamma = (\sigma^x \otimes \mathbf{1}, \sigma^z\otimes \mathbf{1}, \sigma^y\otimes s^x, \sigma^y\otimes s^y, \sigma^y\otimes s^z )$ where $\sigma^\alpha$ ($s^\alpha$) are Pauli matrices corresponding to sublattice/orbital (spin) degree of freedom.

The nearest-neighbor hopping of graphene is given by
\begin{equation}
g_1^{(t)} = t\big(1+2\cos{(x)}\cos{(y)}\big)\ , \qquad g_{12}^{(t)} = -2 t \cos{(x)}\sin{(y)}\ .
\end{equation}
The Semenoff mass term is given by
\begin{equation}
g_2^{(M)} = M\ .
\end{equation}
The Rashba-spin orbit coupling is given by
\begin{eqnarray}
&g_3^{(R)} = \lambda_R ( 1 - \cos{(x)}\cos{(y)} )\ , \qquad &g_4^{(R)} = -\lambda_R \sqrt{3} \sin{(x)} \cos{(y)}\ ,\\
& g_{23}^{(R)} = -\lambda_R \cos{(x)} \sin{(y)}\ , \qquad &g_{24}^{(R)} = \lambda_R \sqrt{3} \sin{(x)}\sin{(y)}\ .
\end{eqnarray}
The Kane-Kele SOC is given by
\begin{equation}
g_{15}^{(\rm KM)} = t_2 (2 \sin{(2x)} -4\sin{(x)}\cos{(y)})\ .
\end{equation}
The SOC in the SI model is given by
\begin{eqnarray}
& g_{13}^{(\rm SI)} = 2 \tilde t_2 \sin{(x+y)}\ , \qquad g_{14}^{(\rm SI)} = 2 \tilde t_2 \sin{(x-y)}\ ,&\\
& g_{15}^{(\rm SI)} = -2 \tilde t_2 \sin{(2x)}\ .&
\end{eqnarray}
In the previous equations, $x=\frac{1}{2} k_x a$ and $y=\frac{\sqrt{3}}{2} k_y a$.

\vspace{20pt}

\section*{Bibliography}

\bibliographystyle{/Users/sr/bib/prsty}
\bibliography{ti-qpt}

\begin{thebibliography}{10}
\bibitem{3DBHZ}
In a recent work, a 3D generalization of the {BHZ}-{H}ubbard model has been
  investigated within {D}ynamical {M}ean-{F}eld {T}heory. For sufficiently
  large $U$ an antiferromagnetic phase of {G}-type has been
  identified\,\cite{amaricci-16prb235112}.
\end{thebibliography}

\end{document}